%% file: HIN-18-006_temp.tex
\pdfoutput=1

\documentclass[11pt,twoside,a4paper,cmspaper,final,collab]{cms-tdr}
\def\svnVersion{494004P}\def\cmsCernNoTag{CERN-EP-2018-249}\def\cmsCernDate{\today}\def\cmsMessage{Published in Physical Review Letters as \href{http://dx.doi.org/10.1103/PhysRevLett.122.152001}{\doi{10.1103/PhysRevLett.122.152001.}}}

\begin{document}\cmsNoteHeader{HIN-18-006}

\hyphenation{had-ron-i-za-tion}
\hyphenation{cal-or-i-me-ter}
\hyphenation{de-vices}
\RCS$HeadURL: svn+ssh://svn.cern.ch/reps/tdr2/papers/HIN-18-006/trunk/HIN-18-006.tex $
\RCS$Id: HIN-18-006.tex 492749 2019-03-25 19:47:29Z katatar $
\newlength\cmsFigWidth
\ifthenelse{\boolean{cms@external}}{\setlength\cmsFigWidth{0.49\textwidth}}{\setlength\cmsFigWidth{0.65\textwidth}}
\ifthenelse{\boolean{cms@external}}{\providecommand{\cmsLeft}{upper\xspace}}{\providecommand{\cmsLeft}{left\xspace}}
\ifthenelse{\boolean{cms@external}}{\providecommand{\cmsRight}{lower\xspace}}{\providecommand{\cmsRight}{right\xspace}}
\cmsNoteHeader{HIN-18-006}

\newcommand {\PbPb}  {\ensuremath{\mathrm{PbPb}}\xspace}
\newcommand {\ak}    {anti-\kt\xspace}
\newcommand {\ptg}    {\ensuremath{\pt^\PGg}\xspace}
\newcommand {\ptj}    {\ensuremath{\pt^\text{jet}}\xspace}
\newcommand {\ptt}    {\ensuremath{\pt^\text{trk}}\xspace}
\newcommand {\etag}    {\ensuremath{\eta^\PGg}\xspace}
\newcommand {\etaj}    {\ensuremath{\eta^\text{jet}}\xspace}
\newcommand {\etat}    {\ensuremath{\eta^\text{trk}}\xspace}
\newcommand {\phig}    {\ensuremath{\phi^\PGg}\xspace}
\newcommand {\phij}    {\ensuremath{\phi^\text{jet}}\xspace}
\newcommand {\phit}    {\ensuremath{\phi^\text{trk}}\xspace}
\newcommand {\sieie}     {\ensuremath{\sigma_{\eta\eta}}\xspace}
\newcommand {\deta}     {\ensuremath{\Delta\eta}\xspace}
\newcommand {\dphi}     {\ensuremath{\Delta\phi}\xspace}
\newcommand {\deltar}   {\ensuremath{\delta r}\xspace}
\newcommand {\deltaR}   {\ensuremath{\delta R}\xspace}
\newcommand {\dphijg}   {\ensuremath{\Delta\phi_{\mathrm{j}\PGg}}\xspace}
\newcommand {\rjetshape}      {\ensuremath{r}\xspace}
\newcommand {\ra}     {\ensuremath{r_\mathrm{a}}\xspace}
\newcommand {\rb}     {\ensuremath{r_\mathrm{b}}\xspace}
\newcommand {\rf}     {\ensuremath{r_\mathrm{f}}\xspace}
\newcommand {\rhor}    {\ensuremath{\rho(r)}\xspace}

\title{Jet shapes of isolated photon-tagged jets in \texorpdfstring{\PbPb and \Pp\Pp\ collisions at $\sqrtsNN = 5.02\TeV$}{PbPb and pp collisions at sqrt(s[NN]) = 5.02 TeV}}

\date{\today}

\abstract{The modification of jet shapes in \PbPb collisions, relative to those in \Pp\Pp\ collisions, is studied for jets associated with an isolated photon. The data were collected with the CMS detector at the LHC at a nucleon-nucleon center-of-mass energy of 5.02\TeV. Jet shapes are constructed from charged particles with track transverse momenta (\pt) above 1\GeVc in annuli around the axes of jets with $\ptj > 30\GeVc$ associated with an isolated photon with $\ptg > 60\GeVc$. The jet shape distributions are consistent between peripheral \PbPb and \Pp\Pp\ collisions, but are modified for more central \PbPb collisions. In these central \PbPb events, a larger fraction of the jet momentum is observed at larger distances
from the jet axis compared to \Pp\Pp, reflecting the interaction between the partonic medium created in heavy ion collisions and the traversing partons.}

\hypersetup{
pdfauthor={CMS Collaboration},
pdftitle={Jet shapes for isolated photon-tagged jets in PbPb and pp collisions at sqrt(s[NN]) = 5.02 TeV},
pdfsubject={CMS},
pdfkeywords={CMS, physics, heavy ions, jets, jet shapes}}

\maketitle

The quark-gluon plasma (QGP)~\cite{Karsch:1995sy}, a deconfined state of quarks and gluons, can be created in relativistic heavy ion collisions. It can be probed with energetic partons emerging from initial hard scattering processes in the same collisions. The outgoing partons eventually fragment, and each forms a jet of collimated particles that can be observed experimentally. The interactions of the partons with the medium, and therefore the modification of the resulting jets, can be related to the thermodynamical and transport properties of the traversed medium~\cite{Appel:1985dq,Blaizot:1986ma,Gyulassy:1990ye,Wang:1991xy,Baier:1996sk,Zakharov:1997uu}. To better understand the dynamics of the QGP, it is important to explore the mechanisms by which the partons lose energy to the medium, whether by radiation, scattering off its point-like constituents, or by some other processes~\cite{Gyulassy:2000er, Djordjevic:2003zk, Ovanesyan:2011xy, Wang:2001ifa, Burke:2013yra}.

The CERN LHC collaborations have studied the medium-induced modifications of jets by measuring the jet yield for a given transverse momentum (\pt)~\cite{Aad:2012vca, Abelev:2013kqa, Aad:2014bxa, Adam:2015ewa, Khachatryan:2016jfl} and jet substructure~\cite{Chatrchyan:2012gw, Chatrchyan:2013kwa, Chatrchyan:2014ava, Aad:2014wha, Khachatryan:2015lha, Khachatryan:2016tfj, Khachatryan:2016erx, Acharya:2017goa, Sirunyan:2017bsd, Aaboud:2017bzv, Sirunyan:2018jqr,Acharya:2018uvf}. In these types of jet measurements, there is limited information on the initial energy of the parton, \ie, before its interaction with the medium. On the other hand, by studying jets produced in association with an electroweak boson, such as a photon or a \PZ\ boson, whose \pt can be precisely measured, the initial parent parton \pt can be tightly constrained, as electroweak bosons do not interact strongly with the medium~\cite{Wang:1996yh, Wang:1996pe, Dai:2012am}. At LHC energies, these types of processes have an additional advantage: jets associated with an electroweak boson are dominated by quark jets for $\ptj > 30\GeVc$~\cite{Neufeld:2010fj}, hence providing information specifically on quark energy loss, and therefore constraining the dependence of energy loss on parton (quark or gluon) flavor~\cite{Casalderrey-Solana:2015vaa, Kang:2017xnc}.

The CMS Collaboration has previously measured the azimuthal correlation and momentum imbalance of isolated photon+jet pairs in proton-proton (\Pp\Pp) and lead-lead (\PbPb) collisions at nucleon-nucleon center-of-mass energies of $\sqrtsNN = 2.76$ and 5.02\TeV~\cite{Chatrchyan:2012gt, Sirunyan:2017qhf}, and of {\PZ}+jet pairs at 5.02\TeV~\cite{Sirunyan:2017jic}. More recently, the fragmentation functions of jets tagged with an isolated photon were measured~\cite{Sirunyan:2018qec}. A photon is considered isolated if the total transverse energy of other particles in a cone of fixed radius around its direction is small after taking into account the underlying event (UE) contributions as explained in Refs.~\cite{Khachatryan:2010fm, Sirunyan:2017qhf}. This definition suppresses dijet events in which a high-\pt photon originates from one of the jets, either via collinear fragmentation of a parton (``fragmentation photons") or via decays of neutral mesons (``decay photons"). The results showed that in central \PbPb collisions there is an excess of low-\pt particles and a depletion of high-\pt particles inside the jet cone. The jet fragmentation functions reflect the momentum distribution inside the parton shower in the longitudinal direction, making it highly sensitive to the hadronization process~\cite{Casalderrey-Solana:2015vaa}. A complementary observable for medium-induced modifications that features reduced sensitivity to hadronization is the jet radial momentum density profile, \ie, the jet shape, which is a measure of the component of the momentum transverse to the jet axis~\cite{Vitev:2008rz, Chien:2015hda}. Jet shape measurements so far were done using inclusive jet~\cite{Chatrchyan:2013kwa,Sirunyan:2018jqr} or dijet samples~\cite{Khachatryan:2016tfj}.

This Letter reports the first measurement of the differential jet shape for jets associated with an isolated photon. The differential jet shape \rhor is defined as
\begin{equation}
\rhor = \frac{1}{\deltar} \frac{\sum_\text{jets} \sum\limits_{\ra < \rjetshape < \rb} ( \ptt / \ptj)} {\sum_\text{jets} \sum\limits_{0 < \rjetshape < \rf} ( \ptt / \ptj) },
\label{eq:rho}
\end{equation}
where $\deltar = \rb - \ra$ is the width of the annulus of inner and outer radii \ra and \rb with respect to the jet axis, respectively, \ptt is the \pt of tracks falling within each annulus of the jet with \ptj, and $\rjetshape = \sqrt{\smash[b]{(\etaj - \etat)^2 + (\phij - \phit)^2}}$ is the distance between the track and the jet axis in pseudorapidity ($\eta$) and azimuthal angle ($\phi$) plane. The distribution is normalized such that the integral inside the range $0 < \rjetshape < \rf$ is unity where $\rf = 0.3$. Hence, \rhor gives a measure of how the \pt of a jet is distributed (over charged particles) in a direction transverse to the jet axis. The analysis uses \PbPb and \Pp\Pp\ data at $\sqrtsNN = 5.02\TeV$ collected in 2015, corresponding to integrated luminosities of 404\mubinv and 27.4\pbinv, respectively.

The central feature of the CMS detector is a superconducting solenoid of 6\unit{m} internal diameter, providing a magnetic field of 3.8\unit{T}. Within the solenoid volume are a silicon pixel and strip tracker, a lead tungstate crystal electromagnetic calorimeter (ECAL), and a brass and scintillator hadron calorimeter (HCAL), each composed of a barrel and two endcap sections. Hadron forward (HF) calorimeters extend the coverage up to $\abs{\eta}=5.2$ and are used for event selection. In addition, in the case of \PbPb events, the HF signals are used to determine the degree of overlap (``centrality'') of the two colliding Pb nuclei~\cite{Chatrchyan:2011sx} and the event-by-event $\phi$ angle of
maximum particle density (``event plane'')~\cite{Chatrchyan:2012xq}. A more detailed description of the CMS detector can be found in Ref.~\cite{Chatrchyan:2008zzk}.

The event samples are selected online with a trigger requiring a photon with $\ptg > 40\GeVc$~\cite{Sirunyan:2017qhf, Sirunyan:2018qec}. Additional requirements are applied offline to remove noncollision events such as beam-gas interactions~\cite{Khachatryan:2016odn}. For jets and photons, the reconstruction algorithms, analysis selections and corrections for the energy scale and resolution are the same as in Refs.~\cite{Sirunyan:2017qhf, Sirunyan:2018qec}. For \PbPb collisions, the event centrality is defined as the fraction of the total inelastic hadronic cross section of these collisions at $\sqrtsNN = 5.02\TeV$, starting at 0\% for the most central collisions, and is evaluated as percentiles of the distribution of the energy deposited in the HF calorimeters~\cite{Chatrchyan:2011sx}. Results are presented in four centrality intervals: 0--10, 10--30, 30--50, and 50--100\%.

The photon candidates are restricted to the barrel of the ECAL, $\abs{\etag} < 1.44$, and are required to have $\ptg > 60\GeVc$. The trigger is fully efficient for these requirements. Electron contamination and anomalous signals caused by the interaction of highly ionizing particles with the photodiodes used for the ECAL readout are removed, as described in Ref.~\cite{Chatrchyan:2012vq}. Background from hadronic showers is rejected by requiring that the ratio of the HCAL over ECAL energy inside a cone of radius $\deltaR = \sqrt{\smash[b]{(\deta)^2 + (\dphi)^2}}  = 0.15$ around the photon candidate is smaller than 0.1~\cite{Khachatryan:2010fm,Chatrchyan:2012vq}. Background contributions from fragmentation and decay photons are rejected by imposing the same isolation requirements as in Refs.~\cite{Sirunyan:2017qhf,Chatrchyan:2012vq}: the \pt sum in a cone of radius 0.4 with respect to the centroid of the cluster, not including the \pt of the cluster and after correcting for the UE (only in \PbPb collisions), is required to be less than 1\GeVc. The dominant remaining background is from ECAL showers initiated by isolated neutral mesons, \eg, \PGpz, \Pgh, and $\omega$, decaying into pairs of photons that, because of their small opening angle, are reconstructed as a single photon. Their contribution can be reduced by a factor of $\sim$2 using an upper limit on the shower shape variable \sieie, which is a measure of the width of the ECAL energy cluster distribution in $\eta$ direction~\cite{Chatrchyan:2012vq, Sirunyan:2017qhf}.

The energy of the reconstructed photons is corrected to account for the losses due to material in front of the ECAL and for incomplete shower containment~\cite{Khachatryan:2015iwa}. An additional correction is applied in \PbPb collisions to account for the contribution of the UE formed by soft processes. The corrections are obtained from photon events simulated using the CUETP8M1 tune~\cite{Khachatryan:2015pea} of the \PYTHIA~8.212~\cite{Sjostrand:2014zea} Monte Carlo (MC) event generator. The effect of the \PbPb UE is modeled by embedding the \PYTHIA output in events generated using \HYDJET 1.9~\cite{Lokhtin:2005px}, which is tuned to reproduce global event properties, such as the UE \pt density, charged-hadron multiplicity and \pt distribution. The size of the resulting energy correction for isolated photons varies from 0 to 10\%, depending on the \ptg and the centrality. The CMS detector response for generated events is simulated using \GEANTfour~\cite{geant4}.

Jets are reconstructed from the output of the CMS particle-flow algorithm~\cite{Sirunyan:2017ulk}, which aims to reconstruct and identify each individual particle in an event, with an optimized combination of information from the various elements of the detector. The \ak algorithm~\cite{Cacciari:2008gp,Cacciari:2011ma} is used to cluster the resulting particles using a distance parameter $R = 0.3$ chosen to minimize the effects of UE fluctuations. In order to subtract the UE background in \PbPb collisions, an iterative algorithm~\cite{Kodolova:2007hd} is employed~\cite{Chatrchyan:2011sx,Chatrchyan:2012nia,Chatrchyan:2012gt}. In \Pp\Pp\ collisions, where the UE level is negligible, jets are reconstructed without UE subtraction. Additional \PbPb (\Pp\Pp) collisions in the same or adjacent bunch crossings are negligible (small) and their effects are found, using MC studies, to be negligible. The jet energy corrections are derived from simulation, separately for \Pp\Pp\ and \PbPb collisions. They are validated via energy balance methods applied to dijet and photon+jet events in \Pp\Pp\ data~\cite{Khachatryan:2016kdb}, reconstructed alternatively with the \Pp\Pp\ and \PbPb reconstruction algorithms. Jets with $\abs{\etaj} < 1.6$ and corrected $\ptj > 30\GeVc$ are selected.

In each event, photon+jet pairs are formed by associating the highest \ptg isolated photon candidate with all jets that pass the jet selection criteria. An azimuthal separation of $\dphijg = \abs{\phij - \phig} > 7\pi/8$ is applied to the photon+jet pairs to suppress contributions from background jets (jets not originating from the same hard scattering as the photon) and from photon+multijet events (the hard scattering produces more than one parton balancing the photon). The tracks used in this measurement have $\ptt>1\GeVc$, $\abs{\etat}<2.4$, and must fall within a cone of radius $\deltaR = 0.3$ around the jet direction. These selection criteria, as well as the corrections for tracking efficiency, detector acceptance, and misreconstruction rate, are the same as in Ref.~\cite{Khachatryan:2016odn} for both \Pp\Pp\ and \PbPb data.

To isolate the contribution of photons, jets, and charged particles that are produced in the same hard scattering in \PbPb collisions, several sources of combinatorial backgrounds are subtracted: tracks from the UE that fall within the cone around the selected jet, misidentified jets resulting from UE fluctuations, and jets not produced in the same hard parton-parton scatterings as the photon. The shape and magnitude of these contributions to the \rhor distributions are estimated from data with an event mixing procedure, in which either the isolated photon or the jet are combined with jets and tracks found in events chosen randomly from a minimum bias (MB) \PbPb data set with similar event characteristics (centrality, interaction vertex position, and event plane angle, which is correlated to particle density in the $\phi$ direction). The background contribution from UE tracks is estimated by constructing the distribution for each selected jet using tracks from MB events. The backgrounds from jets produced by UE fluctuations or a different hard parton-parton scattering are estimated as in Refs.~\cite{Chatrchyan:2012gt,Sirunyan:2017qhf}. The normalizations of these combinatorial background distributions are given by the number of MB events used. Simulation shows that the UE particle density can be different between a hard scattering event (\sloppy{\PYTHIA{}+\HYDJET} ) and a MB event (\HYDJET only) that have the same reconstructed centrality. Therefore, the normalized background distributions are further scaled with a residual factor to account for this effect before being subtracted from those in photon+jet events.

An additional correction is applied for effects such as detector resolution, particle reconstruction, and UE particles uncorrelated to the true jet. This correction is calculated from the \sloppy{\PYTHIA{}+\HYDJET} (\PYTHIA) sample for the \PbPb (\Pp\Pp) results. The distributions from reconstructed (detector-level) jets are corrected to the ones from true (generator-level) jets as a function of \rjetshape. The correction is calculated in three steps. i) The jet shapes for reconstructed jets using reconstructed tracks are corrected to the ones that use true charged particles. This step accounts for the reconstructed track yield that decreases with the distance between the track and jet axis, an effect resulting from the correlation between track reconstruction efficiency and jet reconstruction. The average corrections for $r<0.2 (r>0.2)$ are $\sim 4(5)$\% for \Pp\Pp\ and $\sim 4(10)$\% for 0--10\% centrality \PbPb results. ii) The jet shapes obtained after the first step are corrected to the ones that use true charged particles from the signal \PYTHIA event. This step accounts for the correlations between the reconstructed jet and tracks from the UE and is applied for \PbPb data only. The average corrections for $r<0.2 (r>0.2)$ are $\sim 10(15)$\% for 0--10\% centrality \PbPb results. iii) The jet shapes obtained after the second step are corrected to the ones for true jets. This last step accounts for the difference between the jet shapes for reconstructed and true jets. The average corrections for $r<0.2 (r>0.2)$ are $\sim 2(3)$\% for \Pp\Pp\ and $\sim 20(35)$\% for 0--10\% centrality \PbPb results. The corrections are calculated in bins of \rjetshape, \ptj, \etaj, \ptt, and centrality. The largest corrections happen at $\rjetshape\approx0.3$ and their average values in the first, second, and third steps for 0--10\% centrality \PbPb (\Pp\Pp) collisions are 15 (6)\%, 20 (0)\%, and 45 (4)\%, respectively. Studies have been done separately for the shapes of quark and gluon jets in order to check if the corrections, which do not take parton flavor into account, cause a bias in the results. The corrections improve the agreement between reconstructed and true jets for both quark and gluon jets in both \PYTHIA and {\PYTHIA{}+\HYDJET} samples.

A final correction accounts for the photon purity, defined to be the fraction of photons within the set of isolated photon candidates that do not originate from hadron decays and that pass the \sieie requirement. This fraction is extracted from the data using a template fit to the \sieie distribution~\cite{Sirunyan:2017qhf,Chatrchyan:2012gt}. The shape of the \rhor distributions from decay photons is estimated by repeating the analysis selecting photons with larger \sieie (wider shower shapes). The purity values (\eg, 0.68 and 0.82 for 0--10\% and 50--100\% \PbPb collisions, respectively) from the shower shape fits are used to adjust the magnitude of this background contribution.

Several sources of systematic uncertainty are considered, including the photon purity, photon isolation, photon energy scale, electron contamination, photon selection efficiency, jet energy scale, jet energy resolution, tracking efficiency, \rjetshape-dependent corrections, and background subtraction. The total uncertainty in each bin is the sum in quadrature of the individual uncertainties. The quoted systematic uncertainties are an average over all \rjetshape bins. In the case of the \PbPb results, uncertainties are reported only for the 0--10\% centrality interval, which has generally the highest uncertainties among all the centrality bins.

To evaluate the systematic uncertainties related to the isolated photons, the same procedures are applied as in Ref.~\cite{Sirunyan:2017qhf}. The uncertainty in the photon purity is evaluated by varying the components of the shower shape template, as in Ref.~\cite{Chatrchyan:2012gt}. The maximum variations with respect to the nominal case are propagated as systematic uncertainties, amounting to 0.6 (0.3)\% for the \PbPb (\Pp\Pp) results. In the following, the uncertainties will continue to be quoted for central \PbPb events first, then for \Pp\Pp\ data. The systematic uncertainties resulting from the experimental isolation criteria for a photon are 1.9 and 0.1\%. The residual data-to-simulation photon energy scale difference after applying the photon energy corrections is also quoted as a systematic uncertainty of 0.7\% for \PbPb data, while it is negligible for \Pp\Pp\ data. The level of electron contamination in the samples before applying the electron rejection criteria is 14\% and reduces to roughly 5\% after the rejection procedure. An uncertainty is evaluated by repeating the analysis without applying the electron rejection criteria, and scaling down the difference in the \rhor distribution to the remaining electron contamination after applying the electron rejection, giving 0.3 and $<$0.1\%. The efficiency in selecting photons has been extracted from simulation as a function of photon \pt and data are corrected for this efficiency. An uncertainty is assigned by comparing the results to the ones obtained with a correction derived by loosening the selection criteria, given 0.2 and $<$0.1\%.

The uncertainties related to the jet energy resolution and jet energy scale are evaluated as in Ref.~\cite{Sirunyan:2017qhf}. When propagated, the uncertainty related to the jet energy scale amounts to 6.9 and 0.8\%, while the energy resolution gives uncertainties of 1.9 and 0.3\%. The uncertainty related to the tracking inefficiency is estimated as the difference in the track reconstruction efficiency between data and simulation, as in Ref.~\cite{Khachatryan:2016odn}. Tracking corrections are varied in a \ptt-dependent way, giving systematic uncertainties of 1.0 and 0.9\%.

Further systematic uncertainties are assigned for the \rjetshape-dependent correction procedure. First, it is observed in MC simulations that the first step of corrections has a remaining disagreement of 2\% at $\rjetshape\approx0.3$ between reconstructed tracks and true charged particles, in both the \Pp\Pp\ and \PbPb cases. Second, the model dependence of the corrections is studied by obtaining the quark and gluon jet shape distributions from MC simulations and fitting them to distributions in data. The extracted templates are varied by the fit uncertainty. The difference between the nominal and varied templates is quoted as systematic uncertainty, amounting to 0.5 (1)\% and 3 (4)\% in the $r<0.2(r>0.2)$ case, for \Pp\Pp\ and \PbPb results, respectively.

For \PbPb collisions a systematic uncertainty for the background subtraction is estimated by combining two independent sources. First, results are obtained using an alternative background subtraction procedure (the so-called $\eta$-reflection method~\cite{Chatrchyan:2014ava}) and compared to the nominal method. Second, nominal results are compared to the ones where the background distributions are not scaled for the UE particle density difference seen in simulation. The combined difference of 3.5\% is assigned as the uncertainty.

\begin{figure*}[hbtp]
  \centering
    \includegraphics[width=0.9\textwidth]{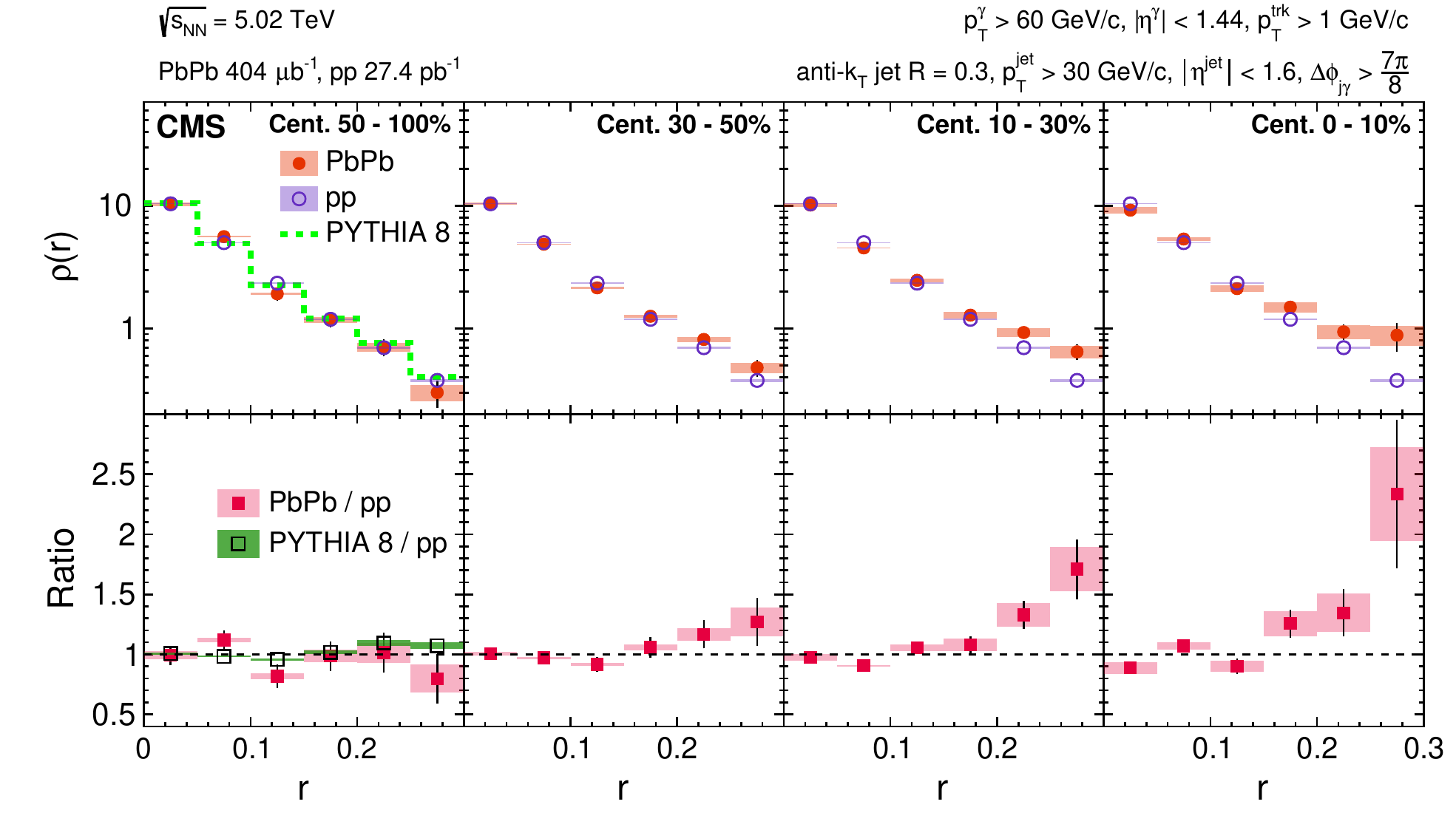}
    \caption{
      Upper:~The differential jet shape \rhor for jets associated with an isolated photon for (from left to right) 50--100\%, 30--50\%, 10--30\%, 0--10\% \PbPb (solid circles), and \Pp\Pp\ (open circles) collisions and from \PYTHIA simulation (histogram). Lower:~The ratios of the \PbPb and \Pp\Pp\ distributions. For the \Pp\Pp\ results, the ratio is to the \PYTHIA distribution. The vertical lines through the points represent statistical uncertainties, while the shaded colored boxes indicate the total systematic uncertainties in data.
      }
      \label{fig:jetshape}
\end{figure*}

The upper panel of Fig.~\ref{fig:jetshape} shows the differential jet shape \rhor for both \PbPb and \Pp\Pp\ collisions, and \PYTHIA simulation. The ratio of \PbPb to \Pp\Pp\ (simulated to \Pp\Pp) data distributions are shown in the lower panel. The simulation is slightly higher than the \Pp\Pp\ data at large \rjetshape, but describes the \Pp\Pp\ data to within 10\% in each bin, allowing its use to derive the \rjetshape-dependent corrections. The uncertainties considered correlated between the \Pp\Pp\ and \PbPb datasets (from photon isolation, photon purity, photon efficiency, electron rejection, jet energy scale, jet energy resolution, tracking efficiency, and from the $r$-dependent procedure corrections) partially cancel in the ratio. The distribution in 50--100\% \PbPb collisions is consistent with that in \Pp\Pp\ collisions. The difference between the \Pp\Pp\ and the 0--10\% (0--30\%) \PbPb results was quantified by comparing the two distributions with a $\chi^2$-test, including all statistical and systematical uncertainties. The $p$-value found was 0.029 (0.017). This shows that, with a $p$-value cutoff of 0.05, the two sets of results are incompatible with each other for the two most central \PbPb collisions bins. In these central collisions, an enhancement of the \rhor distribution with respect to the reference \Pp\Pp\ data is observed at $\rjetshape\approx0.3$. When integrated over different $r$-intervals, the results show that $\sim$5\% of \Pp\Pp\ jet energy is beyond $r > 0.2$. For jets in 0--10\% \PbPb collisions the jet energy fraction changes to $\sim$9\%.  This implies that in \PbPb data a larger fraction of the jet momentum is carried at large distances from the jet axis. The enhancement seen at large \rjetshape is in qualitative agreement with the inclusive jet shape results in Refs.~\cite{Chatrchyan:2013kwa, Sirunyan:2018jqr}, and both the leading and subleading jet shapes in Ref.~\cite{Khachatryan:2016tfj}. In contrast, no significant depletion is seen in central collisions for intermediate \rjetshape, as was observed in the aforementioned inclusive jet shape and leading jet shape results. This could be because of tagging the jet sample with isolated photons, which increases the quark jet fraction, and because of the lower \ptj threshold, which increases the fraction of less collimated jets (including those with a larger relative energy loss). On the other hand, the \rhor distributions decrease rapidly with $r$, with the bulk of the jet energy being concentrated at small $r$ in both collision systems. Since the fraction of \rhor shifted from small to large $r$ because of medium modifications in \PbPb collisions is small compared to the integrated fraction at small $r$, the depletion cannot appear large.

In summary, the differential jet shapes for jets associated with isolated photons are measured in \Pp\Pp\ and \PbPb collisions for the first time. They are constructed using charged particles with transverse momentum $\ptt > 1\GeVc$, for jets with $\ptj > 30\GeVc$, which are associated with an isolated photon with $\ptg > 60\GeVc$. While the distribution from the most peripheral (50--100\%) \PbPb collisions is consistent with that in \Pp\Pp\ data, a modification of the jet shape in \PbPb collisions is observed in more central events. The 0--10\% (10--30\%) \PbPb \rhor is enhanced for the distance between the track and the jet axis $\rjetshape\gtrsim0.15$ (0.20). No significant suppression is seen at intermediate \rjetshape. The modifications demonstrate that for hard scatterings that predominantly produce quarks with similar momentum distributions in \Pp\Pp\ and \PbPb collisions, as identified by the photon tag, the jet momentum is distributed at greater radial distance in \PbPb collisions. This significant redistribution of energy observed in central \PbPb collisions, compared with \Pp\Pp\ and peripheral \PbPb collisions, can be interpreted as a direct observation of jet broadening in the quark-gluon plasma (QGP). This first measurement of radial momentum density profile for jets tagged by an isolated photon, which constrains the information about the jet energy before any loss occurred while traversing the QGP,  constitutes a new unambiguous reference for testing theoretical models of parton-medium interactions.

\begin{acknowledgments}
We congratulate our colleagues in the CERN accelerator departments for the excellent performance of the LHC and thank the technical and administrative staffs at CERN and at other CMS institutes for their contributions to the success of the CMS effort. In addition, we gratefully acknowledge the computing centers and personnel of the Worldwide LHC Computing Grid for delivering so effectively the computing infrastructure essential to our analyses. Finally, we acknowledge the enduring support for the construction and operation of the LHC and the CMS detector provided by the following funding agencies: BMBWF and FWF (Austria); FNRS and FWO (Belgium); CNPq, CAPES, FAPERJ, FAPERGS, and FAPESP (Brazil); MES (Bulgaria); CERN; CAS, MoST, and NSFC (China); COLCIENCIAS (Colombia); MSES and CSF (Croatia); RPF (Cyprus); SENESCYT (Ecuador); MoER, ERC IUT, and ERDF (Estonia); Academy of Finland, MEC, and HIP (Finland); CEA and CNRS/IN2P3 (France); BMBF, DFG, and HGF (Germany); GSRT (Greece); NKFIA (Hungary); DAE and DST (India); IPM (Iran); SFI (Ireland); INFN (Italy); MSIP and NRF (Republic of Korea); MES (Latvia); LAS (Lithuania); MOE and UM (Malaysia); BUAP, CINVESTAV, CONACYT, LNS, SEP, and UASLP-FAI (Mexico); MOS (Montenegro); MBIE (New Zealand); PAEC (Pakistan); MSHE and NSC (Poland); FCT (Portugal); JINR (Dubna); MON, RosAtom, RAS, RFBR, and NRC KI (Russia); MESTD (Serbia); SEIDI, CPAN, PCTI, and FEDER (Spain); MOSTR (Sri Lanka); Swiss Funding Agencies (Switzerland); MST (Taipei); ThEPCenter, IPST, STAR, and NSTDA (Thailand); TUBITAK and TAEK (Turkey); NASU and SFFR (Ukraine); STFC (United Kingdom); DOE and NSF (USA).
\end{acknowledgments}

\bibliography{auto_generated}

\cleardoublepage \appendix\section{The CMS Collaboration \label{app:collab}}\begin{sloppypar}\hyphenpenalty=5000\widowpenalty=500\clubpenalty=5000\input{HIN-18-006-authorlist.tex}\end{sloppypar}
\end{document}

%% file: HIN-18-006-authorlist.tex
\vskip\cmsinstskip
\textbf{Yerevan Physics Institute, Yerevan, Armenia}\\*[0pt]
A.M.~Sirunyan, A.~Tumasyan
\vskip\cmsinstskip
\textbf{Institut f\"{u}r Hochenergiephysik, Wien, Austria}\\*[0pt]
W.~Adam, F.~Ambrogi, E.~Asilar, T.~Bergauer, J.~Brandstetter, M.~Dragicevic, J.~Er\"{o}, A.~Escalante~Del~Valle, M.~Flechl, R.~Fr\"{u}hwirth\cmsAuthorMark{1}, V.M.~Ghete, J.~Hrubec, M.~Jeitler\cmsAuthorMark{1}, N.~Krammer, I.~Kr\"{a}tschmer, D.~Liko, T.~Madlener, I.~Mikulec, N.~Rad, H.~Rohringer, J.~Schieck\cmsAuthorMark{1}, R.~Sch\"{o}fbeck, M.~Spanring, D.~Spitzbart, A.~Taurok, W.~Waltenberger, J.~Wittmann, C.-E.~Wulz\cmsAuthorMark{1}, M.~Zarucki
% ...
\vskip\cmsinstskip
% ...
1:  Also at Vienna University of Technology, Vienna, Austria\\
% ...


%% file: HIN-18-006_temp.bbl
\providecommand{\href}[2]{#2}\begingroup\raggedright\begin{thebibliography}{10}%
\makeatletter
\providecommand{\hrefCMSnoop }[0]{\@secondoftwo}%
\makeatother
\providecommand{\doi}{\texttt{doi:}\begingroup \urlstyle{tt}\Url}

\bibitem{Karsch:1995sy}
\hrefCMSnoop {}{F.~Karsch, ``{The phase transition to the quark gluon plasma:
  recent results from lattice calculations}'',} \textit{ Nucl. Phys. A}
  \textbf{ 590} (1995) 367,
  \href{http://dx.doi.org/10.1016/0375-9474(95)00248-Y}{\doi{10.1016/0375-9474(95)00248-Y}},
\href{http://www.arXiv.org/abs/hep-lat/9503010}{\texttt{arXiv:hep-lat/9503010}}.

\bibitem{Appel:1985dq}
\hrefCMSnoop {}{D.~A. Appel, ``Jets as a probe of quark-gluon plasmas'',}
  \textit{ Phys. Rev. D} \textbf{ 33} (1986) 717,
  \href{http://dx.doi.org/10.1103/PhysRevD.33.717}{\doi{10.1103/PhysRevD.33.717}}.

\bibitem{Blaizot:1986ma}
\hrefCMSnoop {}{J.~P. Blaizot and L.~D. McLerran, ``Jets in expanding
  quark-gluon plasmas'',} \textit{ Phys. Rev. D} \textbf{ 34} (1986) 2739,
\href{http://dx.doi.org/10.1103/PhysRevD.34.2739}{\doi{10.1103/PhysRevD.34.2739}}.

\bibitem{Gyulassy:1990ye}
\hrefCMSnoop {}{M.~Gyulassy and M.~Pl{\"u}mer, ``{Jet quenching in dense
  matter}'',} \textit{ Phys. Lett. B} \textbf{ 243} (1990) 432,
\href{http://dx.doi.org/10.1016/0370-2693(90)91409-5}{\doi{10.1016/0370-2693(90)91409-5}}.

\bibitem{Wang:1991xy}
\hrefCMSnoop {}{X.-N. Wang and M.~Gyulassy, ``Gluon shadowing and jet quenching
  in \textit{A}+\textit{A} collisions at {$\sqrt{s} = 200A$ GeV}'',} \textit{
  Phys. Rev. Lett.} \textbf{ 68} (1992) 1480,
\href{http://dx.doi.org/10.1103/PhysRevLett.68.1480}{\doi{10.1103/PhysRevLett.68.1480}}.

\bibitem{Baier:1996sk}
R.~Baier\hrefCMSnoop {}{ {et~al.}, ``{Radiative energy loss and
  $p_{\perp}$-broadening of high energy partons in nuclei}'',} \textit{ Nucl.
  Phys. B} \textbf{ 484} (1997) 265,
  \href{http://dx.doi.org/10.1016/S0550-3213(96)00581-0}{\doi{10.1016/S0550-3213(96)00581-0}},
\href{http://www.arXiv.org/abs/hep-ph/9608322}{\texttt{arXiv:hep-ph/9608322}}.

\bibitem{Zakharov:1997uu}
\hrefCMSnoop {}{B.~G. Zakharov, ``Radiative energy loss of high-energy quarks
  in finite-size nuclear matter and quark-gluon plasma'',} \textit{ JETP Lett.}
  \textbf{ 65} (1997) 615,
  \href{http://dx.doi.org/10.1134/1.567389}{\doi{10.1134/1.567389}},
\href{http://www.arXiv.org/abs/hep-ph/9704255}{\texttt{arXiv:hep-ph/9704255}}.

\bibitem{Gyulassy:2000er}
\hrefCMSnoop {}{M.~Gyulassy, P.~Levai, and I.~Vitev, ``{Reaction operator
  approach to nonAbelian energy loss}'',} \textit{ Nucl. Phys. B} \textbf{ 594}
  (2001) 371,
  \href{http://dx.doi.org/10.1016/S0550-3213(00)00652-0}{\doi{10.1016/S0550-3213(00)00652-0}},
\href{http://www.arXiv.org/abs/nucl-th/0006010}{\texttt{arXiv:nucl-th/0006010}}.

\bibitem{Djordjevic:2003zk}
\hrefCMSnoop {}{M.~Djordjevic and M.~Gyulassy, ``{Heavy quark radiative energy
  loss in QCD matter}'',} \textit{ Nucl. Phys. A} \textbf{ 733} (2004) 265,
  \href{http://dx.doi.org/10.1016/j.nuclphysa.2003.12.020}{\doi{10.1016/j.nuclphysa.2003.12.020}},
\href{http://www.arXiv.org/abs/nucl-th/0310076}{\texttt{arXiv:nucl-th/0310076}}.

\bibitem{Ovanesyan:2011xy}
\hrefCMSnoop {}{G.~Ovanesyan and I.~Vitev, ``{An effective theory for jet
  propagation in dense QCD matter: jet broadening and medium-induced
  bremsstrahlung}'',} \textit{ JHEP} \textbf{ 06} (2011) 080,
  \href{http://dx.doi.org/10.1007/JHEP06(2011)080}{\doi{10.1007/JHEP06(2011)080}},
\href{http://www.arXiv.org/abs/1103.1074}{\texttt{arXiv:1103.1074}}.

\bibitem{Wang:2001ifa}
\hrefCMSnoop {}{X.-N. Wang and X.-f. Guo, ``{Multiple parton scattering in
  nuclei: Parton energy loss}'',} \textit{ Nucl. Phys. A} \textbf{ 696} (2001)
  788,
  \href{http://dx.doi.org/10.1016/S0375-9474(01)01130-7}{\doi{10.1016/S0375-9474(01)01130-7}},
\href{http://www.arXiv.org/abs/hep-ph/0102230}{\texttt{arXiv:hep-ph/0102230}}.

\bibitem{Burke:2013yra}
\hrefCMSnoop {}{{JET} Collaboration, ``{Extracting the jet transport
  coefficient from jet quenching in high-energy heavy-ion collisions}'',}
  \textit{ Phys. Rev. C} \textbf{ 90} (2014) 014909,
  \href{http://dx.doi.org/10.1103/PhysRevC.90.014909}{\doi{10.1103/PhysRevC.90.014909}},
\href{http://www.arXiv.org/abs/1312.5003}{\texttt{arXiv:1312.5003}}.

\bibitem{Aad:2012vca}
\hrefCMSnoop {}{{ATLAS Collaboration}, ``{Measurement of the jet radius and
  transverse momentum dependence of inclusive jet suppression in lead-lead
  collisions at $\sqrtsNN=2.76\TeV$ with the ATLAS detector}'',} \textit{ Phys.
  Lett. B} \textbf{ 719} (2013) 220,
  \href{http://dx.doi.org/10.1016/j.physletb.2013.01.024}{\doi{10.1016/j.physletb.2013.01.024}},
\href{http://www.arXiv.org/abs/1208.1967}{\texttt{arXiv:1208.1967}}.

\bibitem{Abelev:2013kqa}
\hrefCMSnoop {}{{ALICE Collaboration}, ``{Measurement of charged jet
  suppression in Pb-Pb collisions at $\sqrtsNN = 2.76\TeV$}'',} \textit{ JHEP}
  \textbf{ 03} (2014) 013,
  \href{http://dx.doi.org/10.1007/JHEP03(2014)013}{\doi{10.1007/JHEP03(2014)013}},
\href{http://www.arXiv.org/abs/1311.0633}{\texttt{arXiv:1311.0633}}.

\bibitem{Aad:2014bxa}
\hrefCMSnoop {}{{ATLAS Collaboration}, ``{Measurements of the nuclear
  modification factor for jets in Pb+Pb collisions at $\sqrtsNN=2.76\TeV$ with
  the ATLAS detector}'',} \textit{ Phys. Rev. Lett.} \textbf{ 114} (2015)
  072302,
  \href{http://dx.doi.org/10.1103/PhysRevLett.114.072302}{\doi{10.1103/PhysRevLett.114.072302}},
\href{http://www.arXiv.org/abs/1411.2357}{\texttt{arXiv:1411.2357}}.

\bibitem{Adam:2015ewa}
\hrefCMSnoop {}{{ALICE Collaboration}, ``{Measurement of jet suppression in
  central Pb-Pb collisions at $\sqrtsNN = 2.76\TeV$}'',} \textit{ Phys. Lett.
  B} \textbf{ 746} (2015) 1,
  \href{http://dx.doi.org/10.1016/j.physletb.2015.04.039}{\doi{10.1016/j.physletb.2015.04.039}},
\href{http://www.arXiv.org/abs/1502.01689}{\texttt{arXiv:1502.01689}}.

\bibitem{Khachatryan:2016jfl}
\hrefCMSnoop {}{{CMS Collaboration}, ``{Measurement of inclusive jet cross
  sections in pp and PbPb collisions at $\sqrtsNN=2.76\TeV$}'',} \textit{ Phys.
  Rev. C} \textbf{ 96} (2017) 015202,
  \href{http://dx.doi.org/10.1103/PhysRevC.96.015202}{\doi{10.1103/PhysRevC.96.015202}},
\href{http://www.arXiv.org/abs/1609.05383}{\texttt{arXiv:1609.05383}}.

\bibitem{Chatrchyan:2012gw}
\hrefCMSnoop {}{{CMS Collaboration}, ``{Measurement of jet fragmentation into
  charged particles in pp and PbPb collisions at $\sqrtsNN=2.76\TeV$}'',}
  \textit{ JHEP} \textbf{ 10} (2012) 087,
  \href{http://dx.doi.org/10.1007/JHEP10(2012)087}{\doi{10.1007/JHEP10(2012)087}},
\href{http://www.arXiv.org/abs/1205.5872}{\texttt{arXiv:1205.5872}}.

\bibitem{Chatrchyan:2013kwa}
\hrefCMSnoop {}{{CMS Collaboration}, ``{Modification of jet shapes in PbPb
  collisions at $\sqrtsNN=2.76\TeV$}'',} \textit{ Phys. Lett. B} \textbf{ 730}
  (2014) 243,
  \href{http://dx.doi.org/10.1016/j.physletb.2014.01.042}{\doi{10.1016/j.physletb.2014.01.042}},
\href{http://www.arXiv.org/abs/1310.0878}{\texttt{arXiv:1310.0878}}.

\bibitem{Chatrchyan:2014ava}
\hrefCMSnoop {}{{CMS Collaboration}, ``{Measurement of jet fragmentation in
  PbPb and pp collisions at $\sqrtsNN=2.76\TeV$}'',} \textit{ Phys. Rev. C}
  \textbf{ 90} (2014) 024908,
  \href{http://dx.doi.org/10.1103/PhysRevC.90.024908}{\doi{10.1103/PhysRevC.90.024908}},
\href{http://www.arXiv.org/abs/1406.0932}{\texttt{arXiv:1406.0932}}.

\bibitem{Aad:2014wha}
\hrefCMSnoop {}{{ATLAS Collaboration}, ``{Measurement of inclusive jet
  charged-particle fragmentation functions in Pb+Pb collisions at
  $\sqrtsNN=2.76\TeV$ with the ATLAS detector}'',} \textit{ Phys. Lett. B}
  \textbf{ 739} (2014) 320,
  \href{http://dx.doi.org/10.1016/j.physletb.2014.10.065}{\doi{10.1016/j.physletb.2014.10.065}},
\href{http://www.arXiv.org/abs/1406.2979}{\texttt{arXiv:1406.2979}}.

\bibitem{Khachatryan:2015lha}
\hrefCMSnoop {}{{CMS Collaboration}, ``{Measurement of transverse momentum
  relative to dijet systems in PbPb and pp collisions at
  $\sqrtsNN=2.76\TeV$}'',} \textit{ JHEP} \textbf{ 01} (2016) 006,
  \href{http://dx.doi.org/10.1007/JHEP01(2016)006}{\doi{10.1007/JHEP01(2016)006}},
\href{http://www.arXiv.org/abs/1509.09029}{\texttt{arXiv:1509.09029}}.

\bibitem{Khachatryan:2016tfj}
\hrefCMSnoop {}{{CMS Collaboration}, ``{Decomposing transverse momentum balance
  contributions for quenched jets in PbPb collisions at $\sqrtsNN=2.76\TeV$
  }'',} \textit{ JHEP} \textbf{ 11} (2016) 055,
  \href{http://dx.doi.org/10.1007/JHEP11(2016)055}{\doi{10.1007/JHEP11(2016)055}},
\href{http://www.arXiv.org/abs/1609.02466}{\texttt{arXiv:1609.02466}}.

\bibitem{Khachatryan:2016erx}
\hrefCMSnoop {}{{CMS Collaboration}, ``{Correlations between jets and charged
  particles in PbPb and pp collisions at $\sqrtsNN=2.76\TeV$}'',} \textit{
  JHEP} \textbf{ 02} (2016) 156,
  \href{http://dx.doi.org/10.1007/JHEP02(2016)156}{\doi{10.1007/JHEP02(2016)156}},
\href{http://www.arXiv.org/abs/1601.00079}{\texttt{arXiv:1601.00079}}.

\bibitem{Acharya:2017goa}
\hrefCMSnoop {}{{ALICE Collaboration}, ``{First measurement of jet mass in
  Pb-Pb and p-Pb collisions at the LHC}'',} \textit{ Phys. Lett. B} \textbf{
  776} (2018) 249,
  \href{http://dx.doi.org/10.1016/j.physletb.2017.11.044}{\doi{10.1016/j.physletb.2017.11.044}},
\href{http://www.arXiv.org/abs/1702.00804}{\texttt{arXiv:1702.00804}}.

\bibitem{Sirunyan:2017bsd}
\hrefCMSnoop {}{{CMS Collaboration}, ``{Measurement of the splitting function
  in pp and Pb-Pb Collisions at $\sqrtsNN=5.02\TeV$}'',} \textit{ Phys. Rev.
  Lett.} \textbf{ 120} (2018) 142302,
  \href{http://dx.doi.org/10.1103/PhysRevLett.120.142302}{\doi{10.1103/PhysRevLett.120.142302}},
\href{http://www.arXiv.org/abs/1708.09429}{\texttt{arXiv:1708.09429}}.

\bibitem{Aaboud:2017bzv}
\hrefCMSnoop {}{{ATLAS Collaboration}, ``{Measurement of jet fragmentation in
  Pb+Pb and pp collisions at $\sqrtsNN = 2.76\TeV$ with the ATLAS detector at
  the LHC}'',} \textit{ Eur. Phys. J. C} \textbf{ 77} (2017) 379,
  \href{http://dx.doi.org/10.1140/epjc/s10052-017-4915-5}{\doi{10.1140/epjc/s10052-017-4915-5}},
\href{http://www.arXiv.org/abs/1702.00674}{\texttt{arXiv:1702.00674}}.

\bibitem{Sirunyan:2018jqr}
\hrefCMSnoop {}{{CMS Collaboration}, ``{Jet properties in PbPb and pp
  collisions at $\sqrtsNN=5.02\TeV$}'',} \textit{ JHEP} \textbf{ 05} (2018)
  006,
  \href{http://dx.doi.org/10.1007/JHEP05(2018)006}{\doi{10.1007/JHEP05(2018)006}},
\href{http://www.arXiv.org/abs/1803.00042}{\texttt{arXiv:1803.00042}}.

\bibitem{Acharya:2018uvf}
\hrefCMSnoop {}{{ALICE Collaboration}, ``{Medium modification of the shape of
  small-radius jets in central Pb-Pb collisions at $\sqrt{s_{\mathrm {NN}}} =
  2.76\,\rm{TeV}$}'',} \textit{ JHEP} \textbf{ 10} (2018) 139,
  \href{http://dx.doi.org/10.1007/JHEP10(2018)139}{\doi{10.1007/JHEP10(2018)139}},
\href{http://www.arXiv.org/abs/1807.06854}{\texttt{arXiv:1807.06854}}.

\bibitem{Wang:1996yh}
\hrefCMSnoop {}{X.-N. Wang, Z.~Huang, and I.~Sarcevic, ``{Jet quenching in the
  opposite direction of a tagged photon in high-energy heavy ion
  collisions}'',} \textit{ Phys. Rev. Lett.} \textbf{ 77} (1996) 231,
  \href{http://dx.doi.org/10.1103/PhysRevLett.77.231}{\doi{10.1103/PhysRevLett.77.231}},
\href{http://www.arXiv.org/abs/hep-ph/9605213}{\texttt{arXiv:hep-ph/9605213}}.

\bibitem{Wang:1996pe}
\hrefCMSnoop {}{X.-N. Wang and Z.~Huang, ``{Medium-induced parton energy loss
  in $\ensuremath{\gamma}$+jet events of high-energy heavy-ion collisions}'',}
  \textit{ Phys. Rev. C} \textbf{ 55} (1997) 3047,
  \href{http://dx.doi.org/10.1103/PhysRevC.55.3047}{\doi{10.1103/PhysRevC.55.3047}},
\href{http://www.arXiv.org/abs/hep-ph/9701227}{\texttt{arXiv:hep-ph/9701227}}.

\bibitem{Dai:2012am}
\hrefCMSnoop {}{W.~Dai, I.~Vitev, and B.-W. Zhang, ``{Momentum imbalance of
  isolated photon-tagged jet production at RHIC and LHC}'',} \textit{ Phys.
  Rev. Lett.} \textbf{ 110} (2013) 142001,
  \href{http://dx.doi.org/10.1103/PhysRevLett.110.142001}{\doi{10.1103/PhysRevLett.110.142001}},
\href{http://www.arXiv.org/abs/1207.5177}{\texttt{arXiv:1207.5177}}.

\bibitem{Neufeld:2010fj}
\hrefCMSnoop {}{R.~B. Neufeld, I.~Vitev, and B.~W. Zhang, ``Physics of
  {$Z^0/\gamma^*$}-tagged jets at energies available at the {CERN large hadron
  collider}'',} \textit{ Phys. Rev. C} \textbf{ 83} (2011) 034902,
  \href{http://dx.doi.org/10.1103/PhysRevC.83.034902}{\doi{10.1103/PhysRevC.83.034902}},
\href{http://www.arXiv.org/abs/1006.2389}{\texttt{arXiv:1006.2389}}.

\bibitem{Casalderrey-Solana:2015vaa}
J.~Casalderrey-Solana\hrefCMSnoop {}{ {et~al.}, ``{Predictions for boson-jet
  observables and fragmentation function ratios from a hybrid strong/weak
  coupling model for jet quenching}'',} \textit{ JHEP} \textbf{ 03} (2016) 053,
  \href{http://dx.doi.org/10.1007/JHEP03(2016)053}{\doi{10.1007/JHEP03(2016)053}},
\href{http://www.arXiv.org/abs/1508.00815}{\texttt{arXiv:1508.00815}}.

\bibitem{Kang:2017xnc}
\hrefCMSnoop {}{Z.-B. Kang, I.~Vitev, and H.~Xing, ``{Vector-boson-tagged jet
  production in heavy ion collisions at energies available at the CERN large
  hadron collider}'',} \textit{ Phys. Rev. C} \textbf{ 96} (2017) 014912,
  \href{http://dx.doi.org/10.1103/PhysRevC.96.014912}{\doi{10.1103/PhysRevC.96.014912}},
\href{http://www.arXiv.org/abs/1702.07276}{\texttt{arXiv:1702.07276}}.

\bibitem{Chatrchyan:2012gt}
\hrefCMSnoop {}{{CMS Collaboration}, ``{Studies of jet quenching using
  isolated-photon+jet correlations in PbPb and pp collisions at
  $\sqrtsNN=2.76\TeV$}'',} \textit{ Phys. Lett. B} \textbf{ 718} (2013) 773,
  \href{http://dx.doi.org/10.1016/j.physletb.2012.11.003}{\doi{10.1016/j.physletb.2012.11.003}},
\href{http://www.arXiv.org/abs/1205.0206}{\texttt{arXiv:1205.0206}}.

\bibitem{Sirunyan:2017qhf}
\hrefCMSnoop {}{{CMS Collaboration}, ``{Study of jet quenching with
  isolated-photon+jet correlations in PbPb and pp collisions at $\sqrtsNN =
  5.02\TeV$}'',} \textit{ Phys. Lett. B} \textbf{ 785} (2018) 14,
  \href{http://dx.doi.org/10.1016/j.physletb.2018.07.061}{\doi{10.1016/j.physletb.2018.07.061}},
\href{http://www.arXiv.org/abs/1711.09738}{\texttt{arXiv:1711.09738}}.

\bibitem{Sirunyan:2017jic}
\hrefCMSnoop {}{{CMS Collaboration}, ``{Study of jet quenching with {\PZ}+jet
  correlations in Pb-Pb and pp collisions at $\sqrtsNN=5.02\TeV$}'',} \textit{
  Phys. Rev. Lett.} \textbf{ 119} (2017) 082301,
  \href{http://dx.doi.org/10.1103/PhysRevLett.119.082301}{\doi{10.1103/PhysRevLett.119.082301}},
\href{http://www.arXiv.org/abs/1702.01060}{\texttt{arXiv:1702.01060}}.

\bibitem{Sirunyan:2018qec}
\hrefCMSnoop {}{{CMS Collaboration}, ``{Observation of medium induced
  modifications of jet fragmentation in PbPb collisions using
  isolated-photon-tagged jets}'',} (2018).
  \href{http://www.arXiv.org/abs/1801.04895}{\texttt{arXiv:1801.04895}}.
Submitted to \textit{Phys. Rev. Lett.}

\bibitem{Khachatryan:2010fm}
\hrefCMSnoop {}{{CMS Collaboration}, ``{Measurement of the isolated prompt
  photon production cross section in pp collisions at $\sqrt{s} = 7\TeV$}'',}
  \textit{ Phys. Rev. Lett.} \textbf{ 106} (2011) 082001,
  \href{http://dx.doi.org/10.1103/PhysRevLett.106.082001}{\doi{10.1103/PhysRevLett.106.082001}},
\href{http://www.arXiv.org/abs/1012.0799}{\texttt{arXiv:1012.0799}}.

\bibitem{Vitev:2008rz}
\hrefCMSnoop {}{I.~Vitev, S.~Wicks, and B.-W. Zhang, ``{A Theory of jet shapes
  and cross sections: From hadrons to nuclei}'',} \textit{ JHEP} \textbf{ 11}
  (2008) 093,
  \href{http://dx.doi.org/10.1088/1126-6708/2008/11/093}{\doi{10.1088/1126-6708/2008/11/093}},
\href{http://www.arXiv.org/abs/0810.2807}{\texttt{arXiv:0810.2807}}.

\bibitem{Chien:2015hda}
\hrefCMSnoop {}{Y.-T. Chien and I.~Vitev, ``{Towards the understanding of jet
  shapes and cross sections in heavy ion collisions using soft-collinear
  effective theory}'',} \textit{ JHEP} \textbf{ 05} (2016) 023,
  \href{http://dx.doi.org/10.1007/JHEP05(2016)023}{\doi{10.1007/JHEP05(2016)023}},
\href{http://www.arXiv.org/abs/1509.07257}{\texttt{arXiv:1509.07257}}.

\bibitem{Chatrchyan:2011sx}
\hrefCMSnoop {}{{CMS Collaboration}, ``{Observation and studies of jet
  quenching in PbPb collisions at $\sqrtsNN = 2.76\TeV$}'',} \textit{ Phys.
  Rev. C} \textbf{ 84} (2011) 024906,
  \href{http://dx.doi.org/10.1103/PhysRevC.84.024906}{\doi{10.1103/PhysRevC.84.024906}},
\href{http://www.arXiv.org/abs/1102.1957}{\texttt{arXiv:1102.1957}}.

\bibitem{Chatrchyan:2012xq}
\hrefCMSnoop {}{{CMS Collaboration}, ``{Azimuthal anisotropy of charged
  particles at high transverse momenta in PbPb collisions at
  $\sqrtsNN=2.76\TeV$}'',} \textit{ Phys. Rev. Lett.} \textbf{ 109} (2012)
  022301,
  \href{http://dx.doi.org/10.1103/PhysRevLett.109.022301}{\doi{10.1103/PhysRevLett.109.022301}},
\href{http://www.arXiv.org/abs/1204.1850}{\texttt{arXiv:1204.1850}}.

\bibitem{Chatrchyan:2008zzk}
\hrefCMSnoop {}{{CMS Collaboration}, ``The {CMS} experiment at the {CERN}
  {LHC}'',} \textit{ JINST} \textbf{ 3} (2008) S08004,
\href{http://dx.doi.org/10.1088/1748-0221/3/08/S08004}{\doi{10.1088/1748-0221/3/08/S08004}}.

\bibitem{Khachatryan:2016odn}
\hrefCMSnoop {}{{CMS Collaboration}, ``{Charged-particle nuclear modification
  factors in PbPb and pPb collisions at $\sqrtsNN=5.02\TeV$}'',} \textit{ JHEP}
  \textbf{ 04} (2017) 039,
  \href{http://dx.doi.org/10.1007/JHEP04(2017)039}{\doi{10.1007/JHEP04(2017)039}},
\href{http://www.arXiv.org/abs/1611.01664}{\texttt{arXiv:1611.01664}}.

\bibitem{Chatrchyan:2012vq}
\hrefCMSnoop {}{{CMS Collaboration}, ``{Measurement of isolated photon
  production in pp and PbPb collisions at $\sqrtsNN = 2.76\TeV$}'',} \textit{
  Phys. Lett. B} \textbf{ 710} (2012) 256,
  \href{http://dx.doi.org/10.1016/j.physletb.2012.02.077}{\doi{10.1016/j.physletb.2012.02.077}},
\href{http://www.arXiv.org/abs/1201.3093}{\texttt{arXiv:1201.3093}}.

\bibitem{Khachatryan:2015iwa}
\hrefCMSnoop {}{{CMS Collaboration}, ``{Performance of photon reconstruction
  and identification with the CMS detector in proton-proton collisions at
  $\sqrt{s} = 8\TeV$}'',} \textit{ JINST} \textbf{ 10} (2015) P08010,
  \href{http://dx.doi.org/10.1088/1748-0221/10/08/P08010}{\doi{10.1088/1748-0221/10/08/P08010}},
\href{http://www.arXiv.org/abs/1502.02702}{\texttt{arXiv:1502.02702}}.

\bibitem{Khachatryan:2015pea}
\hrefCMSnoop {}{{CMS Collaboration}, ``{Event generator tunes obtained from
  underlying event and multiparton scattering measurements}'',} \textit{ Eur.
  Phys. J. C} \textbf{ 76} (2016) 155,
  \href{http://dx.doi.org/10.1140/epjc/s10052-016-3988-x}{\doi{10.1140/epjc/s10052-016-3988-x}},
\href{http://www.arXiv.org/abs/1512.00815}{\texttt{arXiv:1512.00815}}.

\bibitem{Sjostrand:2014zea}
T.~Sj{\"o}strand\hrefCMSnoop {}{ {et~al.}, ``{An Introduction to PYTHIA
  8.2}'',} \textit{ Comput. Phys. Commun.} \textbf{ 191} (2015) 159,
  \href{http://dx.doi.org/10.1016/j.cpc.2015.01.024}{\doi{10.1016/j.cpc.2015.01.024}},
\href{http://www.arXiv.org/abs/1410.3012}{\texttt{arXiv:1410.3012}}.

\bibitem{Lokhtin:2005px}
\hrefCMSnoop {}{I.~P. Lokhtin and A.~M. Snigirev, ``{A model of jet quenching
  in ultrarelativistic heavy ion collisions and high-$p_{\mathrm{T}}$ hadron
  spectra at RHIC}'',} \textit{ Eur. Phys. J. C} \textbf{ 45} (2006) 211,
  \href{http://dx.doi.org/10.1140/epjc/s2005-02426-3}{\doi{10.1140/epjc/s2005-02426-3}},
\href{http://www.arXiv.org/abs/hep-ph/0506189}{\texttt{arXiv:hep-ph/0506189}}.

\bibitem{geant4}
\hrefCMSnoop {}{{GEANT4} Collaboration, ``{\GEANTfour}---a simulation
  toolkit'',} \textit{ Nucl. Instrum. Meth. A} \textbf{ 506} (2003) 250,
  \href{http://dx.doi.org/10.1016/S0168-9002(03)01368-8}{\doi{10.1016/S0168-9002(03)01368-8}}.

\bibitem{Sirunyan:2017ulk}
\hrefCMSnoop {}{{CMS Collaboration}, ``Particle-flow reconstruction and global
  event description with the {CMS} detector'',} \textit{ JINST} \textbf{ 12}
  (2017) P10003,
  \href{http://dx.doi.org/10.1088/1748-0221/12/10/P10003}{\doi{10.1088/1748-0221/12/10/P10003}},
\href{http://www.arXiv.org/abs/1706.04965}{\texttt{arXiv:1706.04965}}.

\bibitem{Cacciari:2008gp}
\hrefCMSnoop {}{M.~Cacciari, G.~P. Salam, and G.~Soyez, ``{The anti-\kt jet
  clustering algorithm}'',} \textit{ JHEP} \textbf{ 04} (2008) 063,
  \href{http://dx.doi.org/10.1088/1126-6708/2008/04/063}{\doi{10.1088/1126-6708/2008/04/063}},
\href{http://www.arXiv.org/abs/0802.1189}{\texttt{arXiv:0802.1189}}.

\bibitem{Cacciari:2011ma}
\hrefCMSnoop {}{M.~Cacciari, G.~P. Salam, and G.~Soyez, ``{FastJet user
  manual}'',} \textit{ Eur. Phys. J. C} \textbf{ 72} (2012) 1896,
  \href{http://dx.doi.org/10.1140/epjc/s10052-012-1896-2}{\doi{10.1140/epjc/s10052-012-1896-2}},
\href{http://www.arXiv.org/abs/1111.6097}{\texttt{arXiv:1111.6097}}.

\bibitem{Kodolova:2007hd}
\hrefCMSnoop {}{O.~Kodolova, I.~Vardanian, A.~Nikitenko, and A.~Oulianov,
  ``{The performance of the jet identification and reconstruction in heavy ions
  collisions with CMS detector}'',} \textit{ Eur. Phys. J. C} \textbf{ 50}
  (2007) 117,
\href{http://dx.doi.org/10.1140/epjc/s10052-007-0223-9}{\doi{10.1140/epjc/s10052-007-0223-9}}.

\bibitem{Chatrchyan:2012nia}
\hrefCMSnoop {}{{CMS Collaboration}, ``{Jet momentum dependence of jet
  quenching in PbPb collisions at $\sqrtsNN=2.76\TeV$}'',} \textit{ Phys. Lett.
  B} \textbf{ 712} (2012) 176,
  \href{http://dx.doi.org/10.1016/j.physletb.2012.04.058}{\doi{10.1016/j.physletb.2012.04.058}},
\href{http://www.arXiv.org/abs/1202.5022}{\texttt{arXiv:1202.5022}}.

\bibitem{Khachatryan:2016kdb}
\hrefCMSnoop {}{{CMS Collaboration}, ``{Jet energy scale and resolution in the
  CMS experiment in pp collisions at 8 TeV}'',} \textit{ JINST} \textbf{ 12}
  (2017) P02014,
  \href{http://dx.doi.org/10.1088/1748-0221/12/02/P02014}{\doi{10.1088/1748-0221/12/02/P02014}},
\href{http://www.arXiv.org/abs/1607.03663}{\texttt{arXiv:1607.03663}}.

\end{thebibliography}\endgroup
